\documentstyle[psfig]{lamuphys}
\makeatletter
\let\chapter\hid@chapter
\makeatother
\def\ref{\par\noindent\hangindent=1truecm}
\font\piedi=cmr8

\catcode`\@=11
\def\gsim{\ifmmode{\mathrel{\mathpalette\@versim>}}
    \else{$\mathrel{\mathpalette\@versim>}$}\fi}
\def\lsim{\ifmmode{\mathrel{\mathpalette\@versim<}}
    \else{$\mathrel{\mathpalette\@versim<}$}\fi}
\def\@versim#1#2{\lower 2.9truept \vbox{\baselineskip 0pt \lineskip 
    0.5truept \ialign{$\m@th#1\hfil##\hfil$\crcr#2\crcr\sim\crcr}}}
\catcode`\@=12

\def\lsun{\hbox{$L_\odot$}}
\def\lb{\hbox{$L_{\rm B}$}}
\def\msun{\hbox{$M_\odot$}}
\def\IMLR{Fe$M/L$}
\def\micm{\hbox{$M_{\rm ICM}$}}
\def\mfecm{\hbox{$M_{\rm Fe}^{\rm ICM}$}}

\def\mfes{\hbox{$M_{\rm Fe}^{*}$}}

\def\zfes{\hbox{$Z^{\rm Fe}_{*}$}}
\def\zfecm{\hbox{$Z^{\rm Fe}_{\rm ICM}$}}
\def\ho{\hbox{$H_\circ$}}
\def\h50{\hbox{$\ho /50$}}
\def\yr-1{\hbox{${\rm yr}^{-1}$}}
\begin{document}
\pagenumbering{arabic}
\title{Chemical Evolution on the Scale of Clusters of Galaxies, and Beyond}

\author{Alvio Renzini\inst{}}

\institute{European Southern Observatory, D-85748 Garching b. M\"unchen,
Germany}

\maketitle

\begin{abstract}
Clusters of galaxies allow a direct estimate of the metallicity and
metal production yield on the largest scale so far. The ratio of the
total iron mass in the ICM to the total optical luminosity of the
cluster (the iron mass-to-light-ratio) is the same for all clusters
which ICM is hotter than $\sim 2$ keV, and the elemental proportions
(i.e. the [$\alpha$/Fe] ratio) appear to be solar. From these
evidences it is argued that both the IMF  as well the relative
contributions of SN types are likely to be universal. Constraints on
the past SN activity in galaxy clusters are then derived, and support
is given to the notion that the average SNIa rate was much higher in
the past, i.e. at least 10 times more than currently observed in local
elliptical-s.

It is also argued that
cluster metallicity ($\sim 1/3$ solar) should be taken as
representative of the low-$z$ universe as a whole.  There is now
compelling evidence that the bulk of stars in cluster
as well as in field ellipticals and bulges formed at high 
redshifts ($z\gsim 3$). Since such stars account for at least $\sim
30\%$ of the baryons now locked into stars, it is argued that at least
$30\%$ of stars and metals formed before $z\simeq 3$. As a
consequence, the
metallicity of the universe at $z=3$ is predicted to be $\sim 1/10$ solar.
This requires the cosmic star formation rate to run at least flat from 
$z\sim 1$ to $\sim 5$, which appears to agree with the most recent
direct
determinations of the star formation rate in Lyman-break galaxies.
\end{abstract}
\section{Introduction}

Clusters of galaxies are the largest entities for which we have
chemical information, hence for which chemical evolution can be
studied.
Moreover, there are reasons to believe that low-redshift clusters are
reasonably fair samples of the nearby universe, as far as global star formation
and ensuing chemical evolution are concerned. Therefore, their study
can  allow us to address in a semiempirical way chemical evolution on
the largest possible scale, that of the universe as a whole. 
In doing so, several interesting constraints can be set on the lower
scales as well, such as individual galaxies.
 
This paper is organized as follows. Section 2 presents the current
evidence for the chemical composition of local clusters of galaxies
at low redshift, for both the intracluster medium (ICM) and for
cluster galaxies. 
In Section 3 the production of iron on the scale of clusters is
discussed, setting requirements on the number of Type Ia and Type 2
supernovae (SN) that are necessary to account for the observed amount
of iron. Current theoretical yields are also checked vis \`a vis the
observational constraints. In Section 4 the current evidence  is
briefly reviewed for
the bulk of stars in galactic spheroid-s (i.e. ellipticals and bulges
alike) being very old, formed at high redshift, no matter whether they
reside in rich clusters or in the low density environment we usually
refer to as the {\it field}. In Section 5 these evidences are used to
set constraints on the past history of star formation and metal production,
hence on the metallicity of the high-$z$ universe.

\section{The Chemistry of Galaxy Clusters}

Theoretical simulations predict  that the baryon fraction of rich
clusters
cannot change appreciably in the course of their evolution (White et
al. 1993). 
We can then expect within a cluster
to find confined in the same place 
all the dark matter, all the baryons, all the galaxies, hence all the
metals, that have participated in the play. Clusters are then good {\it
archives} of their past star formation and metal production history.

Metals in clusters are partly spread through their ICM, 
partly locked into galaxies and stars. The mass
of metals in the 
ISM of galaxies is instead negligible compared to that in the two
other components. ICM abundances can be obtained from X-ray 
observations, while optical observations combined to population
synthesis models provide estimates for the metallicity of the stellar
component of galaxies.

\subsection{The Iron Content of the Intracluster Medium}

The existence of large amounts of iron in the ICM was first predicted
on
purely theoretical grounds (Larson \& Dinerstein 1975). Soon
iron was
actually detected via the iron-K X-ray emission at $\sim 7$ keV
(Mitchell et al. 1976).
Fig. 1 shows the 
iron abundance in the ICM of clusters and groups as a function of ICM
temperature, from a compilation of existing data
(Renzini 1997, hereafter R97). As it will become apparent 
later in this section, perhaps even more interesting than the ICM 
abundance of iron is the quantity called 
the iron-mass-to-light-ratio (\IMLR) of the ICM, which is
defined as the ratio $\mfecm/\lb$ of the total iron mass in
the ICM over the total $B$-band luminosity of the galaxies in the
cluster (cf. Songaila, Cowe, \& Lilly 1990; Ciotti et al. 1991; Arnaud
et al. 1992; Renzini et al. 1993; R97). The \IMLR \ of clusters and
groups is shown in Fig. 2.

\begin{figure}
\vskip-4.5truecm
\centerline{\hskip+5mm\psfig{file=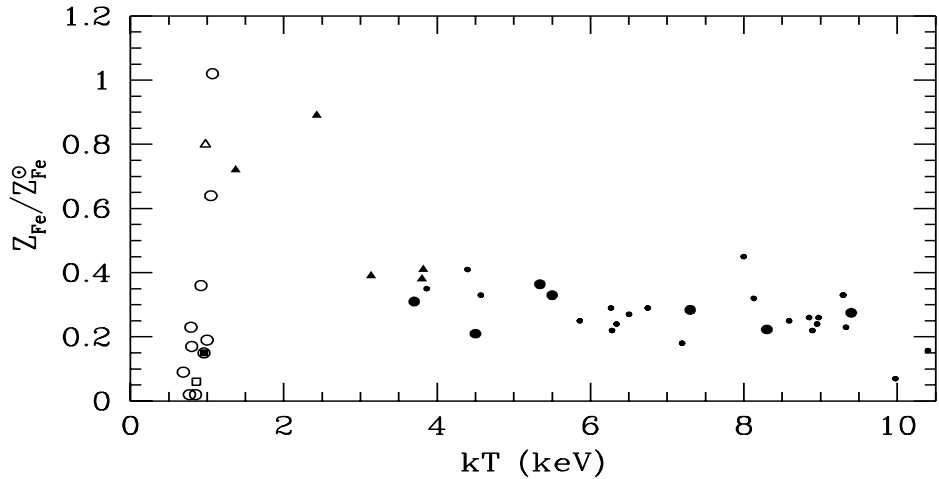,width=13.5cm,angle=0,height=11.2cm}}

\vskip-1truecm
\caption{\piedi A compilation of the iron abundance in the ICM as a 
function of ICM temperature
for a sample of clusters and groups (R97), including several clusters at
moderately high redshift with $<z>\simeq 0.35$, represented by small
filled circles.}
\end{figure}

The drop of the derived \IMLR \ in poor clusters and groups (i.e. for
 $kT\lsim 2$ keV) can be traced back to a drop in both factors
 entering in its definition, i.e., in the iron abundance (cf. Fig. 1)
 {\it and} in the ICM mass to light ratio (R97).  We don't know
 whether this drop is a real effect.  Groups may not behave as closed
 boxes, and may be subject to baryon and metal losses due to strong
 galactic winds driving much of the ICM out of them (Renzini et
 al. 1993; R97: Davis, Mulchaey \& Mushotzky 1998).  
 In addition, there may be a diagnostic problem, since for
 $kT\lsim 2$ keV iron is derived from iron-L transitions involving
 iron ions with 3 to 8 bound electrons.  The atomic physics of iron-L is
 therefore far more complex and uncertain than that of the iron-K,
 which is due to transitions in H-like and He-like iron (Arimoto et al
 1997).  For this reason I will not further discuss clusters whose ICM
 is cooler than $kT\lsim 2$ keV.

Fig. 1 and 2 show that both the iron abundance and the
\IMLR \ in rich clusters ($kT\gsim 2$ keV) are independent of
cluster temperature, hence of cluster richness and
optical luminosity. For these clusters one has 
$\zfecm =0.3\pm 0.1$ solar, and $\mfecm/\lb = (0.02\pm 0.01)$ for
$\ho=50$.
The most straightforward interpretation is that clusters did not lose
iron (hence baryons), nor selectively acquired pristine baryonic material, and
that the conversion of baryonic gas into stars and galaxies has
proceeded with the same efficiency and stellar IMF in all clusters (R97).
Otherwise, we should observe cluster to cluster variations of the iron
abundance and of the \IMLR. 
The theoretical prediction by White et al. (1993) of the constancy of
the baryon fraction in clusters is nicely supported by these evidences. 

\begin{figure}
\vskip-4.5truecm
\centerline{\hskip+5mm\psfig{file=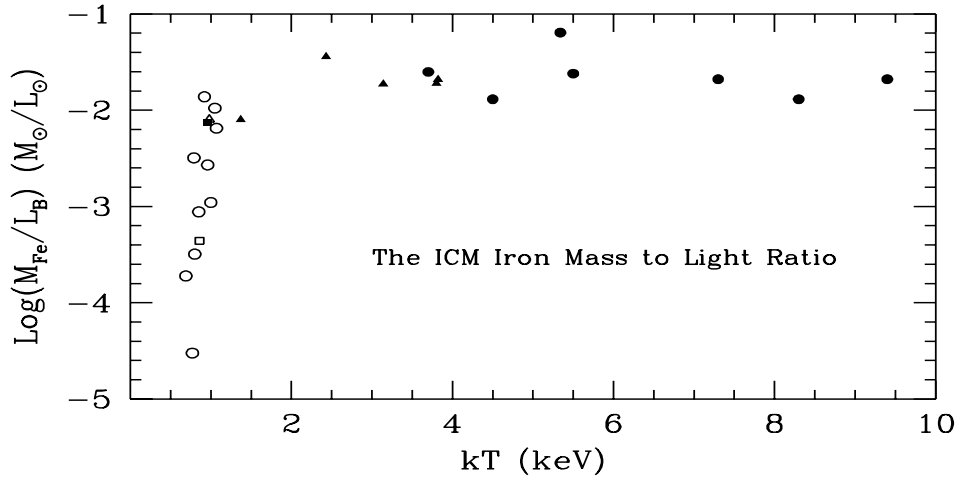,width=13.5cm,angle=0,height=11.2cm}}
\vskip-1truecm
\caption{\piedi The iron mass to light ratio  of  the  ICM of clusters
and groups (for $\ho=50$) as a function of the ICM temperature (R97).}
\end{figure}

\subsection{The $\alpha$-Elements in the Intracluster Medium}

X-ray observations have also allowed to measure the abundance of
other elements in the ICM, especially that of the $\alpha$-elements such as
O, Ne, Mg, and Si, with the {\it ASCA} X-Ray telescope having
superseded all
previous attempts. A high $\alpha$-element
enhancement, 
$<\![\alpha$/Fe]$>\simeq +0.4$, was initially reported (Mushotzky
1994),
an estimate that was soon  revised down to
$<\![\alpha$/Fe]$>\simeq +0.2$  (Mushotzky et al. 1996).
This may still suggest a modest
$\alpha$-element enhancement, with the ICM enrichment being dominated
by SNII products.
However, even this modest $\alpha$-element overabundance vanishes 
when consistently adopting the ``meteoritic'' iron abundance
for the sun, as opposed to the ``photospheric'' value (Ishimaru \&
Arimoto 1997). One can then conclude that
there is no appreciable $\alpha$-element enhancement in
the ICM, with a formal average $<\![\alpha$/Fe]$>\simeq +0.04\;\pm\sim
0.2$ (R97).
Clusters of galaxies as a whole 
are therefore nearly {\it solar} as far as the
elemental ratios are concerned, which argues for stellar 
nucleosynthesis having proceeded in much the same way in the solar 
neighborhood as well as at the galaxy cluster scale. 
In particular, this implies a similar ratio of the number of Type Ia
to Type II SNs, 
as well as
a similar IMF (R97). This result suggests that 
the star formation process (IMF, binary fraction, etc.) is universal, 
with little or no dependence on the global characteristics of the
parent galaxies in which molecular clouds are turned into stars.
Perhaps this is just telling us that like it is difficult to look into
a molecular cloud, it is also difficult for the star forming core of
the cloud to have an idea of the galaxy in which it happens to be
located.

It is nevertheless worth mentioning that other interpretations have
been proposed. For example, bimodal star formation, with most of the
metals in the
ICM having been produced by a now extinct generation of only massive stars, 
has been advocated by Elbaz, Arnaud, \& Vangioni-Flam (1995), see also
Larson (1998). The decoupling of the stellar population having made the
cluster metals, from the one now making the cluster light, would
require fine tuning to account for the observed
constancy of the \IMLR \ which is illustrated in Fig. 2. Such
constancy is instead the straightforward consequence of a universal IMF.

\subsection{The Iron Content of Galaxies and the Iron Share}

The metal abundance of the stellar component of cluster galaxies can
only be inferred from integrated spectra coupled to synthetic stellar 
populations. Much of the stellar mass in clusters is confined to
passively evolving spheroids (ellipticals and bulges) for which the
iron abundance may range from $\sim 1/3$ solar to a few times solar.
Hence, the average iron abundance cannot be much different from solar,
even when taking into account the presence of radial gradients
(Arimoto et al. 1997), $\alpha$-element enhancements (Davies, Sadler,
\& Peletier 1993), and the luminosity bias (Greggio 1997).

The  global iron
abundance of a whole cluster is  therefore given by:
\begin{equation}
Z_{\rm CL}^{\rm Fe}={\zfecm\micm + \zfes M_* \over \micm + M_*}=
      {5.5\zfecm h^{-5/2} + \zfes h^{-1}\over 5.5h^{-5/2} +
      h^{-1}},
\end{equation}
where $\zfes$ is the average abundance of stars in galaxies and $ M_*$
is the mass in stars. For the second equality 
I have assumed as prototypical the Coma cluster values
adopted by White et al. (1993): $\micm\simeq 5.5\times
10^{13}h^{-5/2}\msun$ and $M_*\simeq 10^{13}h^{-1}\msun$.
With $\zfecm=0.3$ solar and $\zfes=1$ solar,
equation (1) gives a global cluster abundance of 0.34,
0.37, and 0.41 times solar, respectively for $h=0.5$, 0.75, and 1. 
Under the same assumptions, the ratio of the iron mass in the ICM to
      the iron mass locked into stars is:
\begin{equation}
{\zfecm\micm\over\zfes M_*}\simeq 1.65 h^{-3/2},
\end{equation}
or 4.6, 2.5, and 1.65, respectively for $h=0.5$, 0.75, and 1. Note
that with the adopted values for the quantities in equation (2) most
of the cluster iron resides in the ICM, rather than being locked into
stars, especially for low values of $\ho$. These estimates could be
somewhat decreased if clusters contain a sizable population of stars
not bound to cense-d individual galaxies, if the average iron abundance
in stars is supersolar (luminosity-weighted determinations
underestimate true abundances, e.g. Greggio 1997), or if the galaxy
$M_*/L$ ratio is larger than the value used here, i.e.,
$<\!M_*/\lb\!>=6.4h$ (White et al. 1993).  In any event, it is clear
that there is at least as much metal mass out of cluster galaxies (in
the ICM), as there is inside them (locked into stars). [Note that the
contribution of the ISM of galaxies is now negligible.]  This must be
taken as a strong constraint by models of the chemical evolution of
galaxies: clearly galaxies do not evolve as a closed box, and outflows
must play a leading role.

With the adopted masses and iron abundances for the two baryonic
components  one can also evaluate the total
cluster \IMLR:
\begin{equation}
{\mfecm +\mfes\over\lb}\simeq 1.3\times 10^{-2}(1.65\, h^{-1/2}+h)
\; (\msun/\lsun),
\end{equation}
or \IMLR=0.037 or 0.034 $\msun/\lsun$, respectively for $h=0.5$ and
1. The total
\IMLR \ is therefore fairly insensitive to the adopted distance scale.

\section{Empirical and Theoretical Metal Yields}

From the near solar proportions of cluster abundances one obtains the
total metal mass to light ratio of a typical cluster as $M_{\rm
Z}/\lb\simeq 10\times M_{\rm Fe}/\lb\simeq 0.3\pm 0.1\;
(\msun/\lsun)$.  It is worth noting that this is an interesting, fully
empirical estimate of the metal yield of stellar
populations. In Section 4 it will be documented that the bulk of stars
in galaxy clusters are very old, say $\sim 15$ Gyr old. Hence, a
single burst approximation may not be too rough for some applications.
This empirical yield  means that $\sim 15$ Gyr after a
burst of star formation there are $\sim 0.3\,\msun$ of metals for each
$\lsun$ of blue light from the surviving (low-mass) stellar
population.
It therefore connects the prompt release of the metals by stars at the
top end of the IMF to the luminosity released  $\sim 15$
Gyr later by the stars at the lower end of the IMF of this same stellar
population. A concrete example may help familiarizing with the
concept. Consider a globular
cluster, with $\lb=10^5\lsun$, $Z=10^{-4}$, age=15 Gyr and
$M=10^5\msun$. With this empirical yield the globular cluster
stellar population has produced $\sim 3\times 10^4\msun$ of metals.
Yet such cluster now contains only $10\,\msun$ of metals, which actually 
pre-existed its formation (as well know the overwhelming majority
of globulars are not self-enriched). All these $\sim 3\times 10^4\msun$
of metals were promptly ejected through a cluster wind driven 
by the $\sim 10^4$ SNIIs that exploded
during the first 30 Myr of the cluster life. Therefore, the cluster was able
to rise to it own metallicity $\sim 3\times 10^8\msun$ of uncontaminated
material, i.e. 3000 times its own present mass. Metal enrichment is a
very quick process!

As in Tinsley (1981), the metal yield is usually
defined per unit mass of stars, a quantity which theoretical
counterpart depends on the poorly known low mass end of the IMF. The
estimate above gives instead the yield per unit luminosity of present
day cluster galaxies, a quantity that depends on the IMF only for
$M\gsim\msun$.  Theoretical mass-related yields have been recently
estimated by Thomas, Greggio, \& Bender (1998) based on massive star
 and supernova explosion models (Woosley \& Weaver 1995; Thielemann,
Nomoto, \& Hashimoto 1996).  These yields can be purged of their
mass dependence, and transformed into luminosity-related yields. 
To this end, let us assume an age of 15 Gyr for the bulk of stars in
clusters (cf. Section 4), and use the proper luminosity-IMF
normalization (Renzini 1998a): i.e. $\psi(M)=AM^{-(1+x)}$ for the IMF
with $A\simeq 3.0\lb$, where $\lb$ is the luminosity of the stellar
population at at the age of 15
Gyr.  Thus, theoretical yields turn out to be $M_{\rm Z}/\lb=0.08$,
0.24, and $0.33\; \msun/\lsun$, respectively for $x=1.7$, 1.35, and
1.00, which compares to $M_{\rm Z}/\lb\simeq 0.3\pm 0.1 \msun/\lsun$
for the empirical cluster value. One can conclude that current stellar
yields do not require a very flat IMF to account for the cluster
metals.

\subsection{The Relative Role of Type Ia and Type II Supernovae}

SN rates are measured in SNUs, with 1 SNU corresponding to $10^{-12}$
SNs $\yr-1 L_{\rm B \odot}^{-1}$. As well known, clusters are now
dominated by E/S0 galaxies, which produce only Type Ia SNs at a rate
of $\sim 0.06$ SNU for $h=0.5$ (Cappellaro et al. 1993). Assuming such rate to
have been constant through cosmological times (15 Gyr), the number of SNIa's
exploded in a cluster of present-day luminosity $\lb$ is therefore
$\sim 6\times 10^{-14}\times 1.5 \times 10^{10}\lb\simeq 10^{-3}\lb$.
With each SNIa producing $\sim 0.7\,\msun$ of iron, the resulting
\IMLR \ of clusters would be:
\begin{equation}
{\left(M_{\rm Fe}\over\lb\right)}_{\rm SNIa}\simeq 7\times 10^{-4},
\end{equation}
which falls short by a factor $\sim 50$ compared to the observed
cluster \IMLR. The straightforward conclusion is that either SNIa's
did not play any significant role in manufacturing iron in clusters, or 
their rate in what are now E/S0 galaxies had to be much higher in the
past. This argues for a strong evolution of the SNIa rate in E/S0
galaxies and bulges, with the past average being some 10 to 50 times
higher than the present rate (Ciotti et al. 1991; R97). 
Observations of SNIa's at high redshift is now becoming a major area of 
observational cosmology (e.g. Garnavich et al. 1998), and whether the
SNIa rate was indeed much higher in the past could soon be tested directly.

In the case of SNIa's we believe to have a fairly precise knowledge of
the amount of iron released by each event, while the ambiguities
affecting the progenitors make theory unable to predict the evolution
of the SNIa rate past a burst of star formation (e.g. Greggio 1996).
The case of Type II SN's is quite the opposite: one believes to have
unambiguously identified the progenitors (stars more massive than
$\sim 8\msun$), while a great uncertainty affects the amount of iron
$M_{\rm Fe}^{\rm II}(M)$ produced by each SNII event as a function of
progenitor's mass. Renzini et al. (1993) list empirical reasons to believe that
this is not a strong function of initial mass, and that assuming
$0.07\,\msun$ of iron per event (as in SN 1987A) cannot be too far
from reality. The total number of SNIIs --$N_{\rm SNII}$-- is obtained
integrating the stellar IMF from e.g. 8 to 100 $\msun$, with the IMF
 being  $\psi(M)=3.0\lb\,M^{-(1+x)},$.

Clearly, the flatter the IMF slope the larger the number of massive stars per
unit present luminosity, the larger the number of SNII's, and therefore the 
larger the implied \IMLR. Thus, integrating the IMF one gets:
\begin{equation}
{\left(M_{\rm Fe}\over\lb\right)}_{\rm SNII}={M_{\rm Fe}^{\rm II}N_{\rm SNII}
         \over\lb}\simeq\cases{
    \hbox{0.003\quad{\rm for}$\; x=1.7$}\hfil\cr
    \hbox{0.009\quad{\rm for}$\; x=1.35$}\hfil\cr
    \hbox{0.035\quad{\rm for}$\; x=0.9.$}\hfil\cr}\
\end{equation}
We conclude that if the Galactic IMF slope ($x=1.7$, Scalo
1986) applies also to ellipticals in rich clusters, then SNII's underproduce
iron by about a factor of 10 compared to what demanded by equation (3). 
Instead, making all the observed iron by SNII's would require a very flat IMF 
($x\simeq 0.9$).

It is worth noting that for an IMF slope
somewhere between Scalo's and Salpeter's SNIIs make about 1/4 to 1/3 of the
total iron, while the rest has to be made by SNIa's, just as in the
so-called {\it standard chemical model} for the chemical evolution of
the Milky Way (cf. Renzini et al. 1993, R97).

\subsection{Clusters as Fair Samples of the Local Universe}

To what extent the cluster global metallicity, and
the ICM to galaxies iron share are representative of the low-$z$
universe as a whole? For example, Madau et al. (1996) adopt $\ho=50$, a stellar
mass density parameter $\Omega_*=0.0036$, and a baryon mass density
parameter $\Omega_{\rm b}=0.05$. With these values the fraction of
baryons that have been locked into stars is $\sim 7\%$. 
This compares to  $\sim
1/(1+5.5h^{-3/2})$ in clusters, or $\sim 6\%$ and $\sim 10\%$,
respectively for $h=0.5$ and 0.75. Fukugita, Hogan \& Peebles estimate
$(\Omega_*/\Omega_{\rm b})_\circ=0.13$ for $h=0.5$, which is a factor of
2 higher than the cluster value, but they may have overestimated the
stellar mass in spheroids (see Section 4.2).
Therefore, it appears that the global, universal efficiency of baryon 
conversion into
galaxies and stars -- $(\Omega_*/\Omega_{\rm b})_\circ$ -- 
is nearly the same as that observed in local clusters, which
supports the notion of clusters being representative of the low-$z$
universe as a whole. 

With nearly the same fraction of baryons having been turned into stars
in the {\it field} as in clusters, one can legitimally entertain also the notion
that no major 
difference in metal enrichment exists between
 field and clusters. Therefore, the global
metallicity of the present-day universe is likely to be $\sim 1/3$ solar,
as that of the only place where we can thoroughly measure it: galaxy
clusters.
If so, the the metal share of the IGM to galaxies should be nearly the
same as the cluster ICM to galaxies metal share, as given by equation (2),
with most of the metals residing in the IGM rather than within field 
galaxies (R97).

\section{The Prompt Initial Enrichment of the Universe}

The cluster abundances as illustrated in the previous section don't say
much about the cosmic epoch when the bulk of the cluster metals were
produced and dispersed through the ICM. The only constraint comes from the
iron abundance in moderate redshift clusters ($z\simeq 0.5$) being
the same of local clusters (see Fig. 1), 
hence the bulk of iron had to be manufactured at $z\gsim 0.5$.
Future X-ray missions could probably bring this limit to $z\simeq 1$.
A much more stringent constraint comes from current age estimates of
the dominant stellar populations in cluster ellipticals, the likely
producers of the bulk of the metals, and from other
{\it fossil} evidences which are concisely reported in this section.
\subsection{The Age of Spheroids}
An important breakthrough on the formation epoch of stars in cluster
ellipticals came from  the very tight 
color-$\sigma$ relation followed by galaxies in the Virgo and Coma
clusters, which demonstrates that the bulk of stars in 
{\it cluster} ellipticals are very old, likely formed at $z\gsim
2$ (Bower, Lucey \& Ellis 1992).
This result had the merit to cut short inconclusive discussions on the
age of ellipticals based on matching synthetic spectra to those of
individual
galaxies, and showed instead that the homogeneity of elliptical
populations sets tight, almost model independent age constraints.
Following the same methodological approach, evidence supporting an
early formation of the bulk of stars in ellipticals
 has greatly expanded over the last few years. This
came from the tightness of the fundamental plane
relation for ellipticals in local clusters (Renzini \& Ciotti 1993),
from the tightness of the color-magnitude relation for ellipticals in
clusters up to $z\sim 1$, and from the
modest shift with increasing redshift in the zero-point of the fundamental
plane, Mg$_2-\sigma$, and color-magnitude relations of cluster
ellipticals (e.g., Aragon-Salamanca et al. 1993; Bender et al. 1997; 
Dickinson 1995; Ellis et
al. 1997; Kodama et al. 1998; Stanford, Esenhardt \& Dickinson 1998;
van Dokkum et al. 1998).
All these studies
agree in concluding that most stars in cluster ellipticals formed at $z\gsim
3$, though the precise value depends on the adopted cosmological
parameters. 

\begin{figure}
\vskip -0.5truecm
\centerline{\hskip+5mm\psfig{file=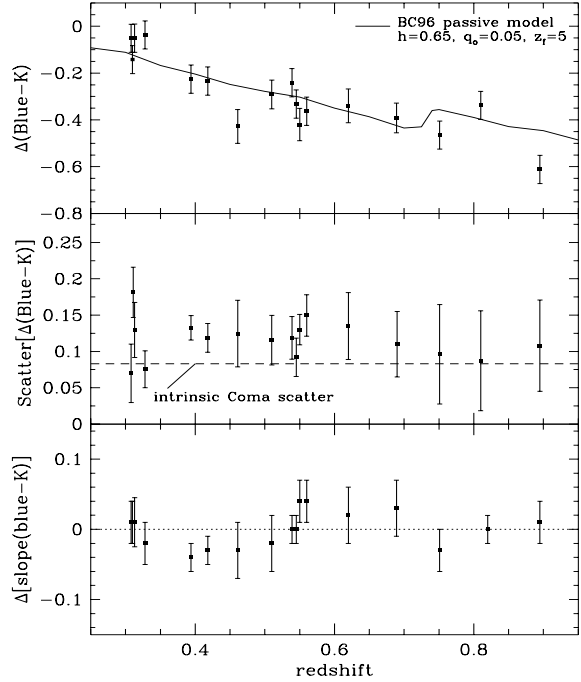,width=8.5cm,angle=0,height=10.2cm}}
\caption[]{The color evolution of early-type galaxies in clusters out
to $z\simeq 0.9$ (Stanford, Eisenhardt, \& Dickinson 1997). 
The ``blue'' band is tuned for each cluster to approximately
sample the rest frame $U$-band, while the $K$ band is always in the
observed frame. Top panel: the redshift evolution of the blue$-K$
color relative to the Coma cluster. A purely passive evolution models
is also shown.  Middle panel: the intrinsic color scatter, having
removed the mean slope of the color-magnitude relation in each cluster
and the contribution of photometric errors. The intrinsic scatter of
Coma galaxies is shown for reference. Bottom panel: the redshift
evolution of the slope of the (blue$-K)-K$ color-mag diagram, modulo
the slope for galaxies in Coma.  }
\end{figure}

This cogent  result has been established for cluster ellipticals, but
additional fossil evidence argues for its validity for field
ellipticals as well as for most bulges of spirals, i.e. in general
for the vast majority of galactic {\it spheroids}.  
Field E/S0 galaxies follow virtually the same 
 Mg$_2-\sigma$ relation of their cluster
counterparts, with the age difference being less than $\sim 1$ Gyr
(Bernardi et al. 1998). Most bulges follow the same
Mg$_2-\sigma$ and color-magnitude relations of ellipticals (Jablonka,
Martin \& Arimoto 1996). 
Even more directly, the bulge of our own
Galaxy is found to be dominated by stars which age is
indistinguishable from that of the Galactic halo, or $\gsim 12$ Gyr
(Ortolani et al. 1995). The case for old spheroidals is more
extensively reviewed 
in Renzini (1999), with an excerpt from Stanford et al. (1998)
being shown in Fig. 3.
\subsection{Demography of Spheroids}
 With spheroids containing at least 30\% of all
stars in the local universe (King \& Ellis 1985; Schechter \& Dressler
1987; Persic \& Salucci 1992), an perhaps as much as $\sim 75\%$
(Fukugita, Hogan \& Peebles 1998), one can rather safely
conclude that at least
 30\% of all stars and metals have formed at $z\gsim 3$ (Renzini
1998b, 1999;
see also Dressler \& Gunn 1990).
This is several times more than suggested by a conservative
 interpretation of the early attempt at tracing the cosmic history of
 star formation, either empirically (Madau et al. 1996) or from 
theoretical simulations (e.g. Baugh et al. 1998). 
Nevertheless, it is in good agreement with more 
recent direct estimates from the spectroscopy of Lyman-break galaxies
(Steidel et al. 1998), where the cosmic SFR runs flat from $z\sim 1$
all the way to $z\sim 5$,
as indeed in one of the plausible options offered by Madau et al. (1998).

On the other hand, the standard CDM model of the Durham group predicts
 that only $\lsim 5\%$ of stars have
formed by $z=3$ (Cole et al. 1994; Baugh et al. 1998). 
However, hierarchical models of galaxy formation may be flexible
 enough to accommodate a major fraction of stars having formed at early
 times. As well known, star formation and its feedback effects are in fact
 incorporated in a quite heuristic fashion into CDM models, adopting simple
 single-parameter algorithms for the rendition of such complex, highly
non-linear phenomena.
\subsection{The Metallicity of the High Redshift Universe}
With $\sim 30\%$ of all stars having formed at $z\gsim 3$, and the
metallicity of the $z=0$ universe being $\sim 1/3$ solar, it is
straightforward
to conclude that the global metallicity of the $z=3$ universe had to
be $\sim 1/3\times 1/3\sim 1/10$ solar,
or more (Renzini 1998b). This opens another area of confrontation,
i.e.
with direct measures of metallicity at high-$z$ that is obtained from
QSO absorbers. The metallicity of DLAs
at $z=3$ appears to be
$\sim 1/20$ solar (Pettini et al. 1997, see their Fig. 4), just a factor of
2  below the predicted value from the {\it fossil evidence}. 
However, the much lower value 
$Z\simeq 10^{-3}Z_\odot$ at $z=3$ has been estimated for the
Ly$_\alpha$ forest (e.g. Songaila 1997). Ly$_\alpha$-forest material
is believed to contain a major fraction of cosmic baryons at high $z$, hence
(perhaps) of metals. There is therefore an extremely large discrepancy
(by perhaps as much as a factor $\sim 300$) with
the estimated global metallicity at $z\simeq 3$ being $\sim 1/10$
solar. This calls into question the notion of 
Ly$_\alpha$ forest metallicity being representative of the the
universe metallicity at this redshift.
Scaling
down from the cluster yield, a metallicity $10^{-3}Z_\odot$ was achieved when
only $\sim 0.3\%$ of stars had formed, which may be largely
insufficient to
ionize the universe at $z>5$ and keep it ionized up to $z=3$ (Madau
1998; see also Gnedin \& Ostriker 1997). 

On the other hand, DLA and Ly$_\alpha$-forest 
systems may provide a vision of the early 
universe that is biased in favor of cold, metal-poor gas that has been
only marginally affected by star formation and metal pollution.
Metal rich objects such as giant starbursts that would be dust obscured, 
the metal rich passively evolving spheroids, and the hot ICM/IGM
obviously do not enlist among QSO absorbers.
This suggests that Ly$_\alpha$ forest may not trace the mass-averaged 
metallicity of high redshift universe, and that the universe was very
inhomogeneous at that epoch. Already at $z\sim3$, the bulk of metals
may be partly locked
into stars in the young spheroidals, partly may reside 
in a yet undetected hot
IGM, a phase hotter than the Ly$_\alpha$ forest phase.
\subsection{Last Speculations}
In conclusion, the fossil evidence on the age of stars in galactic
spheroids coupled to an empirical metal yield of stellar
populations suggests that the universe was already enriched to $\sim 1/10$
solar by $z\simeq 3$. This possible prompt initial enrichment (PIE) of the 
universe may have several interesting ramifications. For example,
it may help explaining the origin of the ubiquitous G-dwarf problem,
such as in the case of the Galactic disk for which a PIE model was
considered very attractive by Tinsley (1981). The idea was that the
stars in the old halo pre-enriched the material than later settled to
form the disk. This
idea can be expanded to say that it was the whole {\it spheroid} (of which the
bulge is by far the major part) promptly enriched to $\sim 1/10$ solar 
a mass some 150 times its own, and that subsequent cooling and infall
of a small fraction of such enriched material gave origin to the
galactic disk.

\smallskip
\noindent
I am grateful to Marc Dickinson for his permission to reproduce Fig. 3
from Stanford et al. (1998).

\end{document}